\documentclass[11pt]{article} 

\usepackage[utf8]{inputenc} 

\usepackage{geometry} 
\usepackage{graphicx} 

\usepackage{booktabs} 
\usepackage{array} 
\usepackage{paralist} 
\usepackage{verbatim} 
\usepackage{subfig} 
\usepackage{graphicx}
\usepackage{multicol}
\usepackage{float}
\usepackage{authblk}

\usepackage{fancyhdr} 
\usepackage{sectsty}
\allsectionsfont{\sffamily\mdseries\upshape} 

\usepackage[nottoc,notlof,notlot]{tocbibind} 
\usepackage[titles,subfigure]{tocloft} 

\title{A Survival Analysis of the Duration of Olympic Records}

\title{A Survival Analysis of the Duration of Olympic Records}
\author[1]{Elliott Hollifield}
\affil[1]{UNC-Asheville\\Asheville, NC}
\author[2]{Victoria Trevi\~no}
\affil[2]{Utah Valley University\\Orem, UT}
\author[3]{Adam Zarn}
\affil[3]{Wheaton College\\Wheaton, IL}
\date{July 26, 2012} 
\begin{document}
\maketitle


\begin{abstract}
\noindent We use recurrent-events survival analysis techniques and methods to analyze the duration of Olympic records. The Kaplan-Meier estimator is used to perform preliminary tests and recurrent event survivor function estimators proposed by Wang \& Chang (1999) and Pena \textit{et al.} (2001) are used to estimate survival curves. Extensions of the Cox Proportional Hazards model are employed as well as a discrete-time logistic model for repeated events to estimate models and quantify parameter significance. The logistic model was the best fit to the data according to the Akaike Information Criterion (AIC). We discuss, in detail, covariate significance for this model and make predictions of how many records will be set at the 2012 Olympic Games in London.\\
\\
\textit{Keywords:} survival analysis, recurrent events, Kaplan-Meier estimator, Cox proportional hazards model, Olympics.

\end{abstract}


\section{Introduction}

Survival analysis is a class of statistical methods for studying the occurrence and timing of events with the ability to incorporate censored observations. Our research is a follow up study of the paper \textit{A recurrent-events survival analysis on the duration of Olympic records} by Guti\'errez, Gonz\'alez, and Lozano (2011), in which survival analysis methods were used to examine the duration of track and field Olympic records. We expand on this previous research by elaborating on censoring as it pertains to this Olympic data, adding many more covariates, and  extending our analysis to canoeing, cycling, and swimming. In terms of modeling, we reproduce previous models estimated, as well as introduce a logistic model that has not yet been applied to Olympic data. \\

As this is a study of time to event data, we defined an event to be a new Olympic record being set. Our main goal was to gain insight about what determines how long Olympic records last. Censoring of a record occurred when we had incomplete information about when a record would have been or will be broken, and these are the two cases of censoring we worked with in this study. The first case is a result of any record that is currently standing. The information about these records is incomplete in the sense that we do not know when current records will be broken. Another case of censoring comes from any record that spanned over the years in which the Olympics were cancelled due to World War I (WWI) or World War II (WWII). The information concerning these records is also incomplete because we do not know if the standing record at that time would have been broken if those particular Olympic Games had not been cancelled.
\section{Data}

Data for this study was collected from \textit{www.olympic.org} and \textit{www.databaseolympics.com}. Given that absolute, as opposed to relative, record times, distances, heights, etc. are necessary for tracking the Olympic record progression of a particular event, the Olympic events analyzed were limited to individual events contained in track \& field, canoeing, cycling, and swimming categories. Record durations span from the first modern Olympic Games held in Athens, Greece in 1896 to the most recent Games held in Beijing, China in 2008. In the total of 27 Olympic Games that have occurred, there have been 690 distinct records in the 63 different Olympic events that were considered. Note that the way in which the event time data was collected did not allow for a record to be both set and broken within the same Olympic Games. Only the time, distance, or height achieved by gold medalists were defined to be the record, if in fact it was better than all prior gold medal performances. This was done to avoid potentially including records which lasted only a very short time (perhaps a few days), which would be assigned a duration of 0 in a discrete model where records can only last a multiple of 4 years.

\subsection{Covariates}

A \textit{covariate} is a factor that has an effect on the length of time that a record lasts. A wide variety of covariates were taken into account in order to characterize a record. We considered many characteristics of specific records and specific Olympic Games. Here is a detailed explanation of the covariates that were considered:

\begin{itemize}

\item {\bf Gender} - 1 for males and 0 for females.\\ \vspace{-.3 in}

\item {\bf Category} - 1 for Track, 2 for Field, 3 for Canoeing, 4 for Cycling, and 5 for Swimming.\\ \vspace{-.3 in}

\item {\bf SameAthlete} - 1 if the previous record was set by the same person, 0 otherwise. \\ \vspace{-.3 in}

\item {\bf SameCountry} - 1 if the previous record was set by a person from the same country, 0 otherwise.\\ \vspace{-.3 in}

\item {\bf HostCountry} - 1 if the record setter is from the same country that the Olympics are being held, 0 otherwise.\\ \vspace{-.3 in}

\item {\bf Medal} - 1 if the person who set the record was an previous Olympic medalist at the time of setting the record, 0 otherwise.\\ \vspace{-.3 in}

\item {\bf WorldRecord} - 1 if the Olympic Record is the World Record, 0 otherwise.\\ \vspace{-.3 in}

\item {\bf Age} - The age of the record setter at the time the record was set.\\ \vspace{-.3 in}

\item {\bf Percent Change Athletes (PCA)} - The percent change of the number of participants.\\ \vspace{-.3 in}

\item {\bf Percent Change in Countries (PCC)} - The percent change in the number of countries participating.\\ \vspace{-.3 in}

\item {\bf dGDP} - Growth rate of Per Capita GDP of the country of the record setter.\\ \vspace{-.3 in}

\item {\bf dPOP} - Growth rate of the population of the country of the record setter.\\ \vspace{-.3 in}

\item {\bf Medal Count (MC)} - The number medals won by the record setter's country through the 2008 Olympics. \\ \vspace{-.3 in}

\item {\bf Number Of Records (NR)} - The  total number of records set at the Olympics in the year a record is set. \\ \vspace{-.3 in}

\item {\bf Marginal Record Improvement (MRI)} - The percent difference between the new record and the previous record.\\ \vspace{-.3 in}

\item {\bf Total Record Improvement (TRI)} - The percent difference between the new record and the first record.\\ \vspace{-.3 in}

\item {\bf World Record Improvement (WRI)} - The percent difference between the Olympic record and the contemporaneous world record.\\

\end{itemize}

Tables 1 - 4 show some simple statistics about our Olympic data. Table~\ref{censored} tells us the number of total observations, number of censored entries, and the number of records that failed, which is how many records have been broken. Table~\ref{freq} tells us the frequency, or number of records, within each Olympic sport category. Table~\ref{covariatefreq} tells us the the number of records that are characterized by either a "1" or a "0" value for the categorical covariates that we included. Table~\ref{Means} gives us some simple summary statistics about our quantitative covariates. Duration, which is the dependent variable, represents the number of years the Olympic record lasted, which has a mean of 6.4 years.

\begin{table}[h]
	\caption{Censored and Uncensored Values}
	\label{censored}
	\centering
	\begin{tabular}{|cccc|}
	\hline
	Total & Failed & Censored & Percent Censored \\
	\hline
	750 & 627 & 123 & 16.40 \\
	\hline
	\end{tabular}
	\centering
\end{table}
\begin{table}[ht]
\begin{tabular}{cc}
	\begin{minipage}[b]{0.5\linewidth}
	\caption{Olympic Category Frequencies}
	\label{freq}
	\centering
	\begin{tabular}{|c|c|c|}
	\hline
	Category & Frequency & Events \\
	\hline
	Track & 218 & 19  \\
	Field & 188 & 13\\
	Canoeing &  37 & 5\\
	Cycling & 18 & 2\\
	Swimming & 289 & 24 \\
	\hline
	Total & 750 & 63\\
	\hline
	\end{tabular}
	\centering
	\end{minipage}
	
&
	\begin{minipage}[b]{0.5\linewidth}
	\caption{Categorical Covariate Frequencies}
	\label{covariatefreq}
	\centering
	\begin{tabular}{|c|c|c|c|}
	\hline
	Variable & N & 1 & 0 \\
	\hline
	Gender & 750 & 518 & 232 \\
	SameAthlete & 627 & 580 & 27 \\
	SameCountry & 628 & 222 & 406 \\
	HostCountry & 688 & 55 & 633 \\
	Medal & 690 & 156 & 534 \\
	WorldRecord & 641 & 243 & 398  \\
	\hline
	\end{tabular}
	\centering
	\end{minipage}
\end{tabular}
\end{table}

\subsection{Dichotomizing Quantitative Covariates}

In order to perform certain analyses described in subsequent sections it was necessary to categorize the quantitave variables in Table~\ref{Means} into distinct groups. In an effort to avoid splitting up these variables in an arbitrary manner, the cutoff was made to be the average value. For example, since the average value for MRI was 2.8, we defined a new variable MRICat as follows:
\[ \textit{MRICat} = \left\lbrace
  \begin{array}{c}
  1, \hspace{2mm} x \geq 2.8  \\ 
   0, \hspace{2mm} x < 2.8 \\
  \end{array} 
  \right.
\]
This kind of dichotomization allowed us to compare record durations with above average values for specific covariates to those with below average values, which would enable us to determine if that covariate has a significant influence on how long the record lasts.

\begin{table}[H]
	\caption{Quantitative Covariate Means\label{Means}}
	\centering
	\begin{tabular}{|c|c|c|c|c|}
	\hline
	Variable & N & Mean & Std Dev & Std Error \\
	\hline
	Duration & 750 & 6.4 & 4.6 & 0.2 \\
	Age & 690 & 22.9 & 4.3 & 0.2 \\
	PCA & 627 & 30.4 & 70.2 & 2.8 \\
	PCC & 689 & 31.0 & 30.1 & 1.1 \\
	dPOP & 645 & 1247.5 & 1627.3 & 64.1 \\
	dGDP & 562 & 247.7 & 277.8 & 11.7 \\
	MC & 690 & 2437.7 & 1908.5 & 72.7\\
	NR & 690 & 30.2 & 9.5 & 0.4 \\
	MRI & 627 & 2.8 & 3.0 & 0.1 \\
	TRI & 690 & 20.5 & 20.7 & 0.8 \\
	WRI & 638 & 1.2 & 1.8 & 0.1 \\
	\hline
	\end{tabular}
	\centering
\end{table}

\section{The Cox Proportional Hazards Model}

The Cox Proportional Hazards Model was used as the basis for some preliminary testing and much of our modeling. Proposed by Sir David Cox (1972), this model is incredibly popular because it does not require one to specify an underlying probability distribution to represent survival times. It is also relatively easy to incorporate time-dependent covariates. The model for the hazard function is written as 
\begin{equation}
\label{Cox}
\lambda(t | X) = \lambda_0(t)e^{\beta_1x_1 + \beta_2x_2 + ... + \beta_px_p}
\end{equation}
where $\lambda_0(t)$ is a baseline hazard function with an unspecified distribution, which is then multiplied by an exponentiated linear function of $p$ fixed covariates. The $\beta$ parameters are estimated by the method of partial likelihood, which Cox developed specifically for this model and is similar to the method of maximum likelihood except that the ordinary likelihood function is treated as the component that depends on $\beta$ alone.  $\lambda_0(t)$ is referred to as the baseline hazard because it is equivalent to $\lambda(t|X=0)$, which is the same as setting all of the covariate values equal to 0. If this were the case we would have no information, but when covariates are included, however, one can generate hazard ratios that give the hazard of one individual, $i$, relative to another individual, $j$:
\begin{equation}
\frac{\lambda_i(t)}{\lambda_j(t)}=e^{\beta_1(x_{i1}-x_{j1})+...+\beta_p(x_{ip}-x_{jp})}.
\end{equation}
Notice that because the baseline is left unspecified and is the same for all individuals, it cancels out from the top and bottom and all that is left is a constant hazard ratio. To give an example, this could be used to determine the relative risks of two records being broken where one is its respective world record and the other is not, but are identical otherwise. The hazard ratio would tell us how much more likely it is for a record that is not the world record to be broken than one that is the world record. In following sections we will use extensions of the Cox Proportional Hazards model to fit our data.

\section{Preliminary Testing}
\subsection{The Kaplan-Meier Estimator}

As mentioned in the previous section, we are interested in estimating the survivor function $S(t)$. If our data contained no censored observations, then the survivor function would be equivalent to the percentage of records still standing at a certain point in time. However, our data does contain censored observations, so the estimation of the survival curve is not as simple, and the best tool for non-recurrent estimation is the Kaplan-Meier (KM) Estimator (1958). Kaplan and Meier proposed that the survivor function could be estimated by
\begin{equation}
\hat{S}(t)=\prod\limits_{j:t_j\le t}\Bigg(1-\frac{d_j}{r_j}\Bigg)
\end{equation}
where $d_j$ is the number of records that were broken at time $t_j$, $r_j$ is the number of records that were at risk of being broken at time $t_j$, and $(1-\frac{d_j}{r_j})$ gives an estimate of the probability of a record lasting to time $t_{j+1}$, given that it has already lasted to time $t_j$. Table~\ref{Example} uses our Olympic data up to time $t_j$=16 to show how KM survivor estimates are calculated.\\  

\begin{table}[h]
\caption{KM Example with Olympic Data\label{Example}}
\centering
\begin{tabular}{|ccc|c|}
\hline
$t_j$ & $d_j$ & $r_j$ & $\hat{S}(t)$ \\ 
\hline \vspace{.1 mm}
$2$   & $3$     & $750$ & $1-\frac{3}{750}=.996$ \\ \vspace{.1 mm}
$4$   & $412$ & $747$ & $.996(1-\frac{412}{747})=.447$ \\ \vspace{.1 mm}
$6$   & $2$     & $256$ & $.447(1-\frac{2}{256})=.443$ \\ \vspace{.1mm}
$8$   & $142$ & $254$ & $.443(1-\frac{142}{256})=.195$ \\ \vspace{.1 mm}
$10$ & $1$     & $95$   & $.195(1-\frac{1}{95})=.193$ \\ \vspace{.1 mm}
$12$ & $43$   & $94$   & $.193(1-\frac{43}{94})=.105$ \\ \vspace{.1 mm}
$16$ & $12$   & $44$   & $.105(1-\frac{12}{44})=.076$ \\
\hline
\end{tabular}  
\centering
\end{table}

Looking at Table~\ref{Example}, one may be wondering how an Olympic record could have lasted two, six, or ten years when Olympic Games are only held every four years. This is a result of the Olympic Games that were held in 1906 to mark the ten year anniversary of the first Olympic Games in 1896. The three records that lasted two years were set in either 1904 or 1906 and broken in 1906 or 1908 respectively, the two that lasted six years were set in 1906 and broken in 1912, and the one record that lasted ten years was set in 1896 and broken in 1906.\\

It must be admitted that with recurrent-event data such as this Olympic data, the KM Estimator is not entirely sufficient for providing theoretically sound survival estimates, the reasons for which will be discussed later in section~\ref{RecurrentEvents}, but we still made use of the KM estimator in our preliminary determination of which covariates to even consider. 

\subsection{Identifying Important Covariates}

After stratifying record durations by categorical covariates that took on values of either 1 or 0, the KM estimator was used to generate two survival curves: one for record durations that had a value of 1 for the covariate of interest, and another that had a value of 0. These two survival curves were then compared to see if there existed a significant difference between them. \\

\begin{figure}[h]
\begin{center}
\includegraphics[scale=.4]{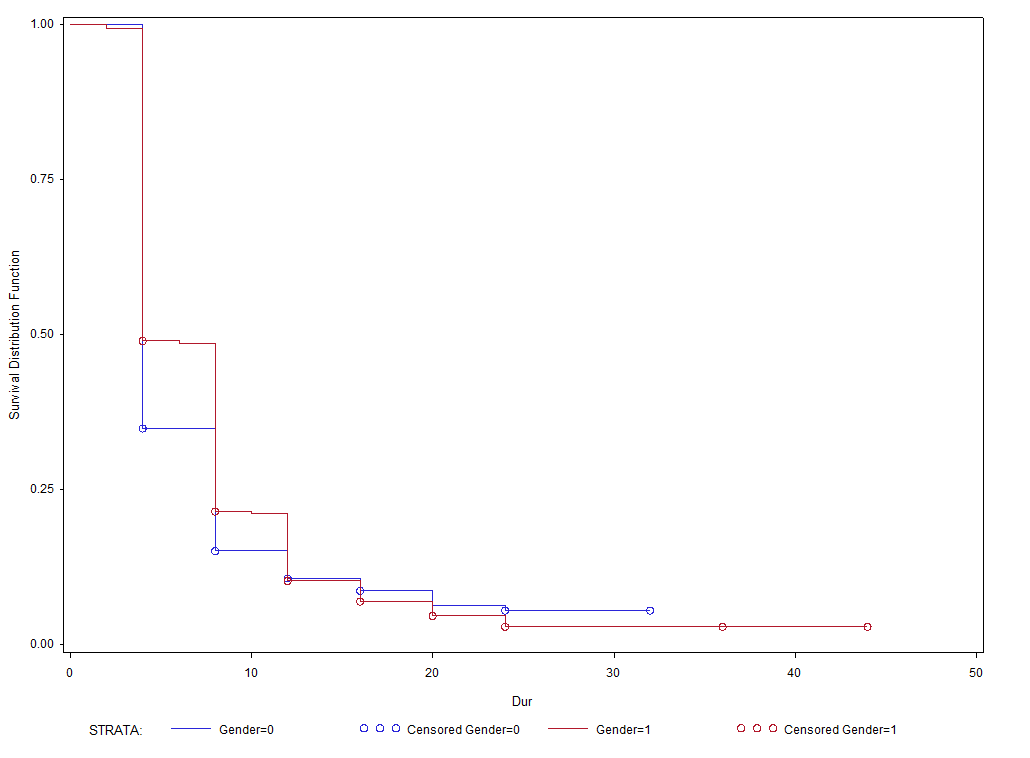}
\end{center}
\caption{KM Survival Curve stratified by Gender\label{Gender}}
\end{figure}

The red and blue curves in Figure~\ref{Gender} represent durations of records set by men and women respectively, and while no definitive decisions can be made regarding the significance of the difference between these two curves from inspecting the graph, the Log-Rank p-value does tell us if there is a significant difference. In the case of Gender, the Log-Rank p-value was .0526, so at the 5 percent level of significance there does not exist enough evidence to reject the null hypothesis of homogeneity of the curves. This led us to believe that there really is no difference between the lengths of records set by men and women, so Gender was excluded from the models we estimated. This same process was applied to each one of the covariates that we considered, and the Log-Rank p-values for each covariate are shown in Table~\ref{significance}. \\

\begin{table}[h]
\caption{Covariate Significance}
\label{significance}
\centering
\begin{tabular}{|c|c || c | c || c | c|}
\hline
Covariate & p-value & Covariate & p-value & Covariate & p-value\\
\hline
Gender&.0526 & HostCountry&.0787 & MCCat & .1503 \\
Category & $<$.0001 & AgeCat & $<$.0001 & NRCat & .2205   \\
SameAthlete & $<$.0001 & PCACat & .0002   &  MRICat & .0003 \\
SameCountry & .0381 & PCCCat & $<$.0001   &  TRICat & .0219  \\
Medal & .0330 &  dGDPCat&.5302 & WRICat&.1714 \\
WorldRecord & .0374 & dPOPCat& .1101 & & \\
\hline
\end{tabular}
\centering
\end{table}

The Covariates that did not demonstrate a significant effect on record durations were Gender, HostCountry, dGDP, dPOP, MC, NR, and WRI. The rest were significant at the 5 percent level and so they remained a part of our analysis and were included in our initial modeling.

\subsection{Dependence Among Observations\label{Dependence}}

An important issue that needed to be addressed was the possibility of dependence among durations of records that were set in the same Olympic event. It is reasonable to expect that two observed durations that came from the same event would be more alike than two randomly selected observations. If this were the case then assumptions of independence would not be met and any techniques we employed that assumed independence would be biased, which would require us to correct for it in some way. To test for dependence, we estimated a Cox model (equation~\ref{Cox}) of the general form 
\begin{equation}
\lambda_{Duration_n}(t | X)  = \lambda_0(t)e^{\beta_0 +\beta_1Duration_{n-k}+ \beta_2x_2 + ... + \beta_px_p}.
\end{equation}
This model estimates the hazard of the $n$th record within an Olympic event being broken, with the length of the $(n-k)$th record as a covariate. All other relevant covariates were also included in the model as we were interested in any residual explanatory power that the $(n-k)$th duration might hold. A significant $\beta$ coefficient on the $(n-k)$th record variable would indicate dependence. We only estimated this model with 1 and 2 lags, corresponding to $k$=1 and $k$=2 in the tables below. We will note here that analyzing higher values of $k$ and including multiple lagged duration variables in one model could certainly be considered.

\begin{table}[H]
\caption{Testing for dependence with $k$=1 and $k$=2\label{dependence}}
\centering
\begin{tabular}{|c|c|c|c || c|c|c|c|}
\hline
\multicolumn{4}{|c| |}{$k$=1} & \multicolumn{4}{c|}{$k$=2} \\
\hline
$n$ & p-value for $\beta_1$ & $n$ & p-value for $\beta_1$ & $n$ & p-value for $\beta_1$ & $n$ & p-value for $\beta_1$ \\
\hline
2 & .9039 & 7 & .7598 & 3 & .5623 & 8 & .8768 \\
3 & .8808 & 8 & .8537 & 4 & .6431 & 9 & .7543\\
4 & .5702 & 9 & .5202 & 5 & .8201 & 10 & .5624\\
5 & .1403 & 10 & .7812 & 6 & .9474 & 11 & .8582\\
6 & .9311 & 11 & .4419 & 7 & .3352 & 12 & .3675\\
\hline
\end{tabular}
\centering
\end{table}

Note that the lowest p-value in Table~\ref{dependence} is .1403 which indicates that if there is some degree of dependence among within-subject observations, it is not significant enough to give us reason to fear bias in procedures that assume independence as a result. However, we thought perhpas there exists greater levels of dependence within certain categories which were not detected when analyzing one unified dataset. This led us to run dependence checks on track and field observations alone and swimming observations alone (canoeing and cycling did not have enough observations to yield meaningful results), the procedures for which were exactly the same as when we did not distinguish by category.

\begin{table}[H]
\caption{Testing for Dependence among Track and Field Observations\label{DepTrackField}}
\centering
\begin{tabular}{|c|c|c|c || c|c|c|c|}
\hline
\multicolumn{4}{|c| |}{$k$=1} & \multicolumn{4}{c|}{$k$=2} \\
\hline
$n$ & p-value for $\beta_1$ & $n$ & p-value for $\beta_1$ & $n$ & p-value for $\beta_1$ & $n$ & p-value for $\beta_1$ \\
\hline
2 & .8370 & 7 & .8588 & 3 & .8173 & 8 & .5286 \\
3 & .6660 & 8 & .8115 & 4 & .3227 & 9 & .9720 \\
4 & .3757 & 9 & .6241 & 5 & .8020 & 10 & .5860 \\
5 & .3140 & 10 & .7695 & 6 & .6952 & 11 & .1826 \\
6 & .9564 & 11 & .4469 & 7 & .1426 & 12 & .7559 \\
\hline
\end{tabular}
\centering
\end{table}
\begin{table}[H]
\caption{Testing for Dependence among Swimming Observations\label{DepSwimming}}
\centering
\begin{tabular}{|c|c|c|c || c|c|c|c|}
\hline
\multicolumn{4}{|c| |}{$k$=1} & \multicolumn{4}{c|}{$k$=2} \\
\hline
$n$ & p-value for $\beta_1$ & $n$ & p-value for $\beta_1$ & $n$ & p-value for $\beta_1$ & $n$ & p-value for $\beta_1$ \\
\hline
2 & .5970 & 7 & .8430 & 3 & .7838 & 8 & .9949 \\
3 & .7535 & 8 & .9949 & 4 & .7625 & 9 & - \\
4 & .7791 & 9 & - & 5 & .8804 & 10 & .9955 \\
5 & .3590 & 10 & - & 6 & .5594 & 11 & - \\
6 & .3371 & 11 & .9964 & 7 & .6121 & 12 & - \\
\hline
\end{tabular}
\centering
\end{table}

As can be seen in Table~\ref{DepTrackField} and Table~\ref{DepSwimming}, there does not appear to be significant dependence among Track and Field observations or Swimming observations. This further supports our conclusion that if dependence does in fact exist among Olympic record durations, there is not enough to bias our results.


\section{Recurrent Event Survival Estimation \label{RecurrentEvents}}

Many of the Kaplan-Meier assumptions are not met with recurrent-event data such as independence of within-subject observations and independence of censoring. The former is not such an obvious problem with this particular Olympic data set, as was discussed in section~\ref{Dependence}, but there is certainly dependence of censoring, as whether or not an observation is censored depends on the time at which the record was set in relation to WWI, WWII, or 2012. So in order to produce accurate survival estimates it was necessary to investigate estimators that have been developed to handle recurrent events.

\subsection{Wang-Chang (1999)}
Wang $\&$ Chang  proposed the following survival probability estimator for recurrent events:

\begin{equation}
\hat{S}_{n} \left( t \right) = \prod\limits_{y^{\ast}_{i} \leq t} \left\lbrace 1 - \frac{d^{\ast} \left( y_{i} \right) }{R^{\ast}\left( y_{i} \right) }\right\rbrace
\end{equation}

\noindent where  $ d^{\ast} $ represents the sum of proportions of individual records set to the total number of records at risk of being broken at time $t$ and $ R^{\ast} $ represents a sum of the average number of times a record has been set for each Olympic sport since time $t$. This estimator assumes within-unit interoccurence times to be correlated.

\subsection{Pe\~na, Strawderman, and Hollander (2001)}
Pe\~na \textit{et al.} (2001) proposed two recurrent event survival functions. One is a generalization of the Kaplan-Meier estimator where the events are recurrent and independently and identically distributed (i.i.d.). In this case the survivor function is estimated by
\begin{equation}
\hat{S}(t)=\prod\limits_{w\le t}\Bigg\{1-\frac{N(\Delta w)}{Y(w)}\Bigg\}
\end{equation}
where, in the period $[0,s]$, N(t) is the number of records whose durations were at most t and Y(t) is the number of records whose durations were at least t.\\

The other, known as the MLE Frailty model, uses a gamma distribution with shape and scale parameters set equal to an unknown parameter $\alpha$ and takes the form
\begin{equation}
\hat{S(t)}= \Bigg[ \frac {\hat{\alpha}} {\hat{\alpha} + \hat{\Lambda}_0(t)} \Bigg] ^ {\hat{\alpha}} ,
\end{equation}
where $\hat{\Lambda}_0(t)$ is an estimator of the marginal cumulative hazard function. The specific value for $\alpha$ is inversely related to the amount of correlation that exists between record duration times; that is, as $\alpha$ increases, correlation decreases, so letting $\alpha \to \infty$ would imply complete independence. \\

\begin{figure}[H]
\centering
\includegraphics[scale=.43]{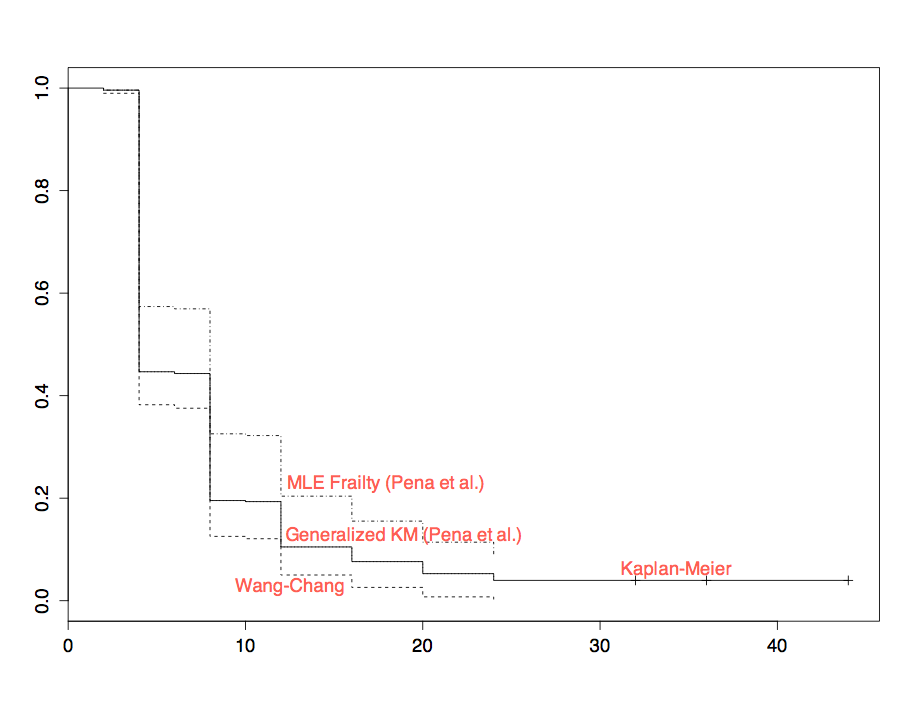}
\vspace{-.35 in}
\caption*{\bf Years}
\caption{Survival Curves developed by Kaplan-Meier, Wang-Chang, and Pe\~na \textit{et al.}}
\label{SurvivalCurves}
\centering
\end{figure}

\begin{table}[H]
\caption{Survival Estimates}
\label{surv_est}
\centering
\begin{tabular}{|c||c|c|c|c|}
\hline
& Kaplan-Meier & Wang-Chang & Generalized KM & MLE Frailty \\
\hline
$\hat{S}(2)$ & .9960  & .9898 & .9960 & .9960 \\
$\hat{S}(4)$ & .4467  & .3823 & .4467 & .5738 \\ 
$\hat{S}(6)$ & .4432  & .3755 & .4432 & .5694 \\
$\hat{S}(8)$ & .1954  & .1255 & .1954 & .3256 \\ 
$\hat{S}(10)$ & .1934 & .1209 & .1934 & .3222 \\
$\hat{S}(12)$ & .1049 & .0501 & .1049 & .2039 \\ 
$\hat{S}(16)$ &.0763  & .0259 & .0763 & .1552 \\
$\hat{S}(20)$ &.0528  & .0075 & .0528 & .1141 \\
\hline
\end{tabular}
\centering
\end{table}

To determine which recurrent event survival estimator produces the most accurate estimates for our Olympic data, the main consideration is dependence. Since the generalized Kaplan-Meier estimator assumes that survival times are independently and identically distributed, it might be the best choice given that our testing showed little signs of dependence. However, these tests were certainly not definitive and perhaps the other two estimators that do allow for dependence would be more appropriate. At the current moment, the 2012 Olympics have not yet occurred, but once those results are in we plan to compare the predictions we make using the different survival estimators in section~\ref{Predictions} with what actually comes to pass in London, which will give us more insight as to which estimator is the most accurate. \\


\section{Modeling}

\subsection{Modeling Recurrent Events with the Cox Model}
The following is a discussion of models that we applied to our Olympic data that extend the Cox Proportional Hazards model to handle recurrent events. Making a choice between these models depends on whether different events have different baseline hazards and if dependence exists among within-event record durations.\\

Instead of assuming that the hazard varies as a function of time since the last event took place, Andersen and Gill (1982) suggested a counting process model that incorporates a hazard that varies as a function of time since the process began. In our case, this specification would imply that record durations depend on the time they were set relative to 1896 as opposed to when the previous record was set. The hazard for the $j$th record in the $i$th Olympic event is then given by 
\begin{equation}
\lambda_{ij}(t|Z_{ij}) = \lambda_0(t)e^{\beta\prime Z_{ij}(t)}
\end{equation}
where $Z_{ij}$ is a vector of covariates. This model is able to take the start and stop times as well as the length of a record into account by specifying a common origin, 1896, for the hazard function. It must be noted that implicit in this model is an assumption of independence between within-event record durations.\\

Prentice \textit{et al.} (1981) proposed two partially-parametric hazard models which call for stratifying data by the sequence time at which a record was broken within an event, so it is essentially a stratifed Andersen-Gill model. The first one depends on the time from the beginning of the study, $t$, and takes the form
\begin{equation}
\lambda\{t|N(t),Z(t)\} = \lambda_{0s}(t)e^{Z(t)\beta_s}
\end{equation}
while the other depends on the time since the immediately preceeding broken record, $t-t_{n(t)}$, which is represented by
\begin{equation}
\lambda\{t|N(t),Z(t)\} = \lambda_{0s}(t-t_{n(t)})e^{Z(t)\beta_s}
\end{equation}
where $N(t) = \{n(u) : u \leq t \}$, where $n(u)$ is the number of records broken prior to time $u$, $Z(t)$ is a vector of covariates, and s is the Olympic event. Essentially, $N(t)$ is a vector of the number of records broken in each event prior to time $t$. These are conditional models in the sense that a record is never at risk of becoming the $n$th record broken within an event until the $(n-1)$th record has been broken, and since observations are stratified by sequence, these models do allow for dependence. \\

Wei \textit{et al.} (1989) developed the most widely used marginal recurrent events model. For the $k$th record of the $i$th Olympic event, the hazard function is given by
\begin{equation}
\lambda_{ki}(t) = \lambda_{k0}(t)e^{{{\beta}^{\prime}}_k Z_{ki}(t)}
\end{equation}
where ${{\beta}^{\prime}}_k = {(\beta_{1k},...,\beta_{pk})}^{\prime}$ is the record-specific regression parameter. This model is basically a combination of the ones already discussed, except that, in regards to the Andersen-Gill model, only the stop times of a record duration are considered and no claim about independence or dependence is made at all. This model is also somewhat controversial given that all records are put into the initial risk set even though a record cannot be broken until the previous one has been broken.

\subsection{Repeated Events for Discrete-Time Maximum Likelihood}

Given that the Olympics only occur once every four years, which restricts records to being broken at only those distinct times, we felt it necessary to consider a discrete model capable of handling repeated events. This forced us to move from the Cox model to a logistic model with slight modification to our data set and inclusion of the time since a record was set as a covariate. \\

If we let $\hat{P}$ depend on $k$ covariates, and also let $\hat{P}$ represent the estimated probability of an individual record $i$ being broken at time $t$, given that record survived to time $t$, we have

\begin{equation}
\hat{P} = \frac{e^{\beta_{k} x_{ik}}}{1+e^{\beta_{k} x_{ik}}}
\end{equation}

\noindent where $\beta_{k}$ is a vector of similar estimated covariate parameters and $x_{ik}$ is a vector of unique covariate values. The logistic model states that the natural logarithm of the odds of an Olympic record being broken is equivalent to a linear combination of $x_{i}$ and $\beta_{k}$:

\begin{equation} 
\log \left( \frac{ \hat{P} }{1- \hat{P}} \right) = \alpha_{t}  +   \beta_{1} x_{i1} + \dots +  \beta_{k} x_{ik}
\end{equation}

\noindent where ${\alpha}_t$ is a constant. \\

In order to use this method we had to modify our data set by creating one entry for each Olympic Games that the record was at risk of being broken. This may be best demonstrated with an example. See below the adjustment made from Table~\ref{uno} to Table~\ref{dos} of the first few entries of an example data set without any explanatory variables. 

\begin{table}[h]
\caption{Example Data}
\label{uno}
\centering
\begin{tabular}{|c|c|c|c|}
\hline 
ID & DUR & CENSOR & SEQ \\ 
\hline 
1 & 4 & 1 & 1 \\ 
\hline 
1 & 8 & 1 & 2 \\ 
\hline 
1 & 8 & 1 & 3 \\ 
\hline 
1 & 12 & 1 & 4 \\ 
\hline 
\end{tabular}
\centering
\end{table}

\begin{table}[h]
\caption{Adjusted Example Data}
\label{dos}
\centering
\begin{tabular}{|c|c|c|c|c|c|}
\hline 
ID & DUR & CENSOR & SEQ & Time & Term \\ 
\hline 
1 & 4 & 1 & 1 & 4 & 1 \\ 
\hline 
1 & 8 & 1 & 2 & 4 & 0 \\ 
\hline 
1 & 8 & 1 & 2 & 8 & 1 \\ 
\hline 
1 & 8 & 1 & 3 & 4 & 0 \\ 
\hline 
1 & 8 & 1 & 3 & 8 & 1 \\ 
\hline 
1  & 12 & 1 & 4 & 4 & 0 \\ 
\hline 
1 & 12 & 1 & 4 & 8 & 0 \\ 
\hline 
1 & 12 & 1 & 4 & 12 & 1 \\ 
\hline 
\end{tabular} 
\centering
\end{table}

\noindent Each entry in our new data set represents a specific Olympic record with new time and term variables. The time variable represents the number of years at $t_{i}$ a record remained unbroken. Term takes on a value of 1 if the record was broken during that specific Olympics, otherwise 0.\\

Unlike proportional hazards models where dependence on time is automatically allowed in the unspecified baseline hazard function, this model includes the time and time-squared since the last event as covariates to allow for a quadratic relationship. Another difference from the Cox model is that, instead of partial likelihood, the method of maximum likelihood is used for parameter estimation.\\


\section{Results \label{Results}}
\subsection{Models}
We began by estimating each model with each one of the covariates we determined to be significant in our preliminary testing. After obtaining results for each of these models, we threw out any covariates that were not significant and estimated the models again. It was the case with every model that all the covariates that were significant in initial modeling maintained significance at the 10 percent level in the subsequent models. Table~\ref{Results} shows our results. Note, we did not include hazard ratios in Table~\ref{Results} for models that extend the Cox model, but they can be obtained by exponentiating e by the parameter estimate. \\

One will notice that while the Medal variable was deemed significant when record durations were stratified by it alone, none of the estimated models found it to be significant in the presence of other significant covariates. Contrastingly, the only variable that every model found to be significant was TRI, while WorldRecord was highly significant in all but the Prentice Gap Time model. The Andersen-Gill and Prentice Elapsed Time models were identical in terms of covariates that were included significantly, but their levels of fit are vastly different, the former being the worst fit and the latter being the best of the Cox models. The marginal model proposed by Wei \textit{et al.} was unique in that it was the only model to retain SameCountry and Age as significant covariates, though it was the second best fit of the Cox models. \\

\begin{figure}[h]
\begin{center}
\includegraphics[scale=.8]{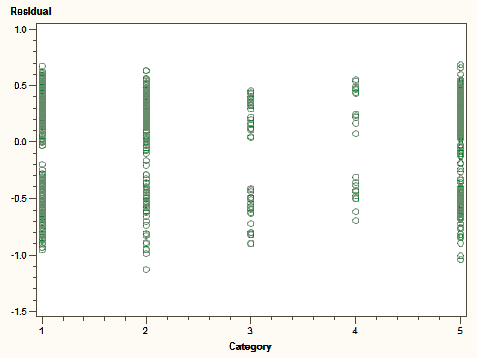}
\end{center}
\caption{Residual Plot by Category for the Logistic Model \label{Residuals}}
\end{figure}

Comparison of the estimated Akaike Information Criterion (AIC) of each model suggests that the discrete time repeated events logistic model is the best fit to the data. This may be explained by the model’s ability to incorporate time dependence. Graphically, as seen in Figure~\ref{Residuals}, the plots of residuals are well behaved for this model which further suggest a good fit. Specifically, Figure~\ref{Residuals} shows the residuals to be approximately normally distributed about zero with similar variance and a lack of outliers.\\

The covariates that this model found to be significant are SameAthlete, WorldRecord, PCC, MRI, and TRI. An interpretation of the parameter estimate for SameAthlete is, holding all other covariates constant, if the previous record, at a given discrete time $t$, was set by the same person who holds the current record, the odds that it will be broken at the next olympics would be 1.3488 units greater than otherwise. Similarly, for WorldRecord, holding all other covariates constant, if the current Olympic record is the world record, then the odds that it will be broken at the next Olympics will be 0.2344 units less than if it were not the world record. Similar interpretations can be used for other parameter estimates.

\begin{table}[H]
\caption{Parameter Estimates\label{Results}}
\begin{center}
\begin{tabular}{|c|c|c|c|c|c|}
\hline
Model & Anderesen-Gill & Prentice & Prentice & Wei & Logistic \\
  & & Elapsed Time & Gap Time & & \\
\hline\hline
Variable  & & & & & \\
\hline
Category & 0.04749*** & 0.05791*** & - & -0.18867*** & \\
 & (0.01135) & (0.01403) & & (0.02961) & \\
SameAthlete & 0.18584*** & 0.21663*** & 0.30032***  & - & 1.3488*** \\
 & (0.04841) & (0.08323) & (0.06835) & & (0.4564) \\
SameCountry & - & - & - &    0.26817***      &            -             \\
                           &                                        &                     &                      &    (0.08612)           &                           \\
Medal                  &                  -                    &  -                 &        -             &           -               &              -          \\
                           &                                        &                     &                       &                           &                           \\
WorldRecord       &          -0.15565***           &  -0.20494***  & -0.08685     &    0.27417***       &    -0.2344**        \\ 
                          &       ( 0.04623)                  &   (0.05416)     & (0.05561)     &    (0.08941)        &      (0.1033)         \\ 
Age                    &                    -                    &         -            &       -            &     -0.03416***    &           -              \\
                          &                                          &                       &                     &   (0.01113)          &                           \\ 
PCA                    &         -0.00303***            &  -0.00375***   &       -             &   0.00298***       &              -           \\
                          &        (0.00104)                  &   (0.00106)      &                     &    (0.0008610)      &                          \\ 
PCC                    &        -0.00515***             & -0.00422***    & -0.00715***  &          -                &      -0.0108***    \\
                          &         (0.00140)                 &  (0.00153)       & (0.00119)      &                            &       (0.00138)      \\
MRI                    &         -                              &          -             &   0.02677***   &   0.09459***      &           0.0549*     \\
                          &                                         &                        &     (0.00985)    &     (0.01429)      &           (0.0313)     \\ 
TRI                     &      0.00321***                 &  0.00848***     &  0.00422***     &   0.02157***     &        0.00915*        \\
                          &      (0.00119)                    &  (0.00233)       &  (0.00161)       &    (0.00403)       &        (0.00553)       \\ 
\hline\hline
AIC                     &    3699.353                     &   1326.748       &    3073.826     &   2625.573         &         892.066 \\
\hline
\end{tabular}
\vspace{.2 in}
\caption*{*, **, and *** denote statistical significance at the 10, 5, and 1 percent levels respectively. Robust sandard errors in parentheses.}
\end{center}
\end{table}

\subsection{Predictions}
\label{Predictions}
We used the survival estimates in Table~\ref{surv_est} to make predictions of how many records we expect to be broken in the 2012 Olympic Games in London. To do this, we needed to know the different probabilities of a record being broken at time $t+4$ given that it has survived to time $t$, for values of $t$ equal to 0, 4, 8, 12, etc. This conditional probability is given by
\begin{equation} \label{predictions}
P(T\le t+4\hspace{1mm}|\hspace{1mm}T > t) = \frac{\hat{S}(t) - \hat{S}(t+4)}{\hat{S}(t)}.
\end{equation}

For example, to find the probability of a record that was set in 2008 being broken in 2012, we would plug 0 in for $t$ in equation \ref{predictions}. 

\begin{table}[H]
\caption{Estimated Number of Records set in 2012 \label{Pred}}
\centering
\begin{tabular}{|c || c|c|c|c|c || c|}
\hline
\multicolumn{1}{|c||}{} & \multicolumn{5}{c||}{Year Set} & \multicolumn{1}{c|}{} \\
\hline
Survival Estimator & 2008 & 2004 & 2000 & 1996 & 1992 & Total \\
\hline
Wang-Chang & 20.38 & 4.03 & 2.40 & 2.90 & 1.42 & 31.14 \\
Generalized KM & 18.26 & 3.38 & 1.85 & 1.64 & .62 & 25.74 \\ 
MLE Frailty & 14.06 & 2.60 & 1.50 & 1.43 & .53 & 20.12 \\
\hline
\end{tabular}
\centering
\end{table}

Since there are so few record durations of more than 20 years, we were unable to produce reliable estimates of the survival function for $t$ greater than 20, as seen in Table~\ref{surv_est}, so only records that were set after 1988 could be used to make predictions. Of the 63 events that were analyzed, the current records for 51 of them were set in 1992 or later, so it was these 51 records that we were able to incorporate into our predictions found in Table~\ref{Pred}. Therefore, of the 51 records included, the Wang-Chang, the Generalized KM, and the MLE Frailty predict that 31.14, 25.74, and 20.12 of them will be broken in 2012 respectively.


\section{Conclusions}

The factors that go in to new Olympic records being set may seem like a mystery to some and completely arbitrary to others. However, we ventured to shed some light on the issue using survival analysis methods and techniques. We first used the Kaplan-Meier estimator to determine which covariates significantly affected the duration of a record and also ran checks for dependence to inform our subsequent analysis. After prleminary measures, we estimated five different models with significant covariates. Four of these models were extensions of the Cox Proportional Hazards model while the other was a discrete-time logistic model for repeated events. In the end, the logistic model was the best fit to our data based on the Akaike Information Criterion, and it depended primarily on the following covariates: SameAthlete, WorldRecord, PCC, MRI, and TRI. We also used survival estimates from three different recurrent event survivor function estimators to determine the number of new records that will be set in the 2012 Olympics. In 51 of the 63 events we considered, the estimated number of records that will be broken is somewhere between 20.12 and 31.14. \\

Future research might include considering more events in the analysis. This could mean adding different individual events, as well as team and winter events. We would also like to investigate covariates that were not included in our analysis such as ethnicity, a measure of technological improvements, and percentage change of males and females competing.


\section{Acknowledgements}
We would like to thank California State University, Fresno and the National Science Foundation for their financial support (NSF Grant \#DMS-1156273), the California State University, Fresno Mathematics REU program, and Dr. Ke Wu for his support during the completion of the project.


\end{document}